\documentclass[aps,prl,twocolumn,superscriptaddress,floatfix,showpacs,nofootinbib]{revtex4-2}
\usepackage{amssymb,amsmath}
\usepackage{graphicx}
\usepackage{subfigure}
\usepackage[english]{babel}
\usepackage{float}
\usepackage{color}
\usepackage[symbol]{footmisc}

\usepackage{graphicx}
\usepackage{natbib}
\usepackage{xcolor}
\usepackage[colorlinks,allcolors=blue]{hyperref}
\usepackage{amsmath}
\usepackage{ulem}
\usepackage{lipsum}

\usepackage{braket}

\newcommand{\eq}[1]{Eq.~(\ref{#1})}

\begin{document}

\title{Spectral edge-to-edge topological state transfer in diamond photonic lattices}

\author{Gabriel C\'aceres-Aravena$^{\dagger}$}
 \affiliation{Departamento de F\'isica, Facultad de Ciencias F\'isicas y Matem\'aticas, Universidad de Chile, Chile}
 \affiliation{Millenium Institute for Research in Optics - MIRO, Chile}
\author{Basti\'an Real$^{\dagger}$}
 \affiliation{Departamento de F\'isica, Facultad de Ciencias F\'isicas y Matem\'aticas, Universidad de Chile, Chile}
 \affiliation{Millenium Institute for Research in Optics - MIRO, Chile}
 \author{Diego Guzm\'an-Silva}
 \affiliation{Departamento de F\'isica, Facultad de Ciencias F\'isicas y Matem\'aticas, Universidad de Chile, Chile}
 \affiliation{Millenium Institute for Research in Optics - MIRO, Chile}
 \author{Paloma Vildoso}
 \affiliation{Departamento de F\'isica, Facultad de Ciencias F\'isicas y Matem\'aticas, Universidad de Chile, Chile}
 \affiliation{Millenium Institute for Research in Optics - MIRO, Chile}
 \author{Ignacio Salinas}
 \affiliation{Departamento de F\'isica, Facultad de Ciencias F\'isicas y Matem\'aticas, Universidad de Chile, Chile}
 \affiliation{Millenium Institute for Research in Optics - MIRO, Chile}
\author{Alberto Amo}
\affiliation{Univ. Lille, CNRS, UMR 8523—PhLAM—Physique des Lasers Atomes et Molécules, F-59000 Lille, France}
\author{Tomoki Ozawa}
\affiliation{Advanced Institute for Materials Research (WPI-AIMR), Tohoku University, Sendai 980-8577, Japan}
\author{Rodrigo A. Vicencio}
\email{rvicencio@uchile.cl}
 \affiliation{Departamento de F\'isica, Facultad de Ciencias F\'isicas y Matem\'aticas, Universidad de Chile, Chile}
 \affiliation{Millenium Institute for Research in Optics - MIRO, Chile}



\begin{abstract}
Transfer of information between topological edge states is a robust way of spatially manipulating quantum states while preserving their coherence in lattice environments. This method is particularly efficient when the edge modes are kept within the topological gap of the lattice during the transfer. In this work we show experimentally the transfer of photonic modes between topological edge states located at opposite ends of a dimerized one-dimensional photonic lattice. We use a diamond lattice of coupled waveguides and show that the transfer is insensitive both to the presence of a high density of states in the form of a flat band at an energy close to that of the edge states, and to the presence of disorder in the hoppings. We explore dynamics in the waveguide lattice using wavelength-scan method, where different input wavelength translates into different effective waveguide length. These results open the way to the implementation of more efficient protocols based on the active driving of the hoppings.

\end{abstract}
\date{\today}

\maketitle
Topological edge states are a remarkable resource to engineer photonic systems with isolated modes protected from the presence of disorder. In two-dimensional lattices, they can be used to fabricate topological edge mode lasers with distributed gain and quantized orbital momentum~\cite{Bahari2017, Bandres2018}, to transfer single photons around corners in elaborated photonic circuits~\cite{Barik2018, Jalali2020}, and to design topological frequency combs with enhanced efficiency~\cite{Mittal2018, mittal_topological_2021}. One dimensional systems such as the Su-Schrieffer-Heeger (SSH) lattice are particularly interesting because topological edge and interface modes are hosted deep into the topological gap of the lattice. This gap protection has been shown to be beneficial to preserve the quantum state of photons in boundary modes~\cite{Blanco-Redondo2018, Tambasco2018}. Interestingly, the presence of topological edge modes on both sides of one-dimensional lattices can be used to transfer a state from one edge of the lattice to the other with high fidelity with the advantage of being protected from certain types of disorder due to the topological nature of the system. Such edge state transfer is a promising route to store and manipulate photonic quantum states in on-chip lattice environments.

Most topological edge transfer protocols rely on the adiabatic evolution of the lattice such that an edge mode is driven into quasi-bulk modes and again into an edge mode at the other side~\cite{Kraus2012, Liu2022, Chapman2016, Lang2017, Li2018, Mei2018, Qi2020, Chen2021, Palaiodimopoulos2021}. While these protocols present an optimized transfer rate and fidelity, they are limited by the adiabaticity condition that requires the adiabatic passage to be slow enough to avoid the Zenner coupling of the edge state information into the bulk modes~\cite{Boross2019, Longhi2019}. Furthermore, the presence of disorder in the lattice would enhance this coupling. A variation of these protocols include counter-adiabatic driving methods~\cite{DAngelis2020}. Recently, a different route has been proposed based on the coherent coupling of edge modes within the gap~\cite{Estarellas2017, Longhi2019, Yuan2021}. The great advantage of this approach is that edge modes are kept well into the topological gap throughout the protocol, ensuring a high fidelity in reduced times. The simplest version of the coherent state transfer of topological edge states is via passive evanescent coupling of the exponential tails of edge modes at opposite sides of the finite size lattice. In this case, coherent transfer between edge modes takes place at a frequency determined by the tail overlap, which can be controlled via the size of the gap. Observation of such coherent oscillations was reported in a short SSH lattice for Rydberg atoms~\cite{DeLeseleuc2019}.

In this work, we demonstrate coherent edge-to-edge transfer of light in a dimerized diamond lattice of coupled waveguides employing a spectral tomographic technique. More importantly, we show experimentally that the fidelity of the transfer is robust to a number of perturbations in the system. First, we show that orthogonality of eigenmodes in our undriven protocol preserves the transfer even in the presence of a high density of states in the form of a flat band at energies close to that of the edge states. Second, we demonstrate that the transfer mechanism is robust to the presence of lattice defects thanks to the underlying chiral symmetry of the system. The experimental proof of principle we report in this work can be significantly sped up by applying a number of driven techniques based on the modulating of the hoppings in time and the use of concatenated topological lattices~\cite{Longhi2019,Yuan2021,Zurita2022}. Such techniques are readily implementable in lattices of coupled waveguides.

\begin{figure}[t!]
\centering
\includegraphics[width=0.99\columnwidth]{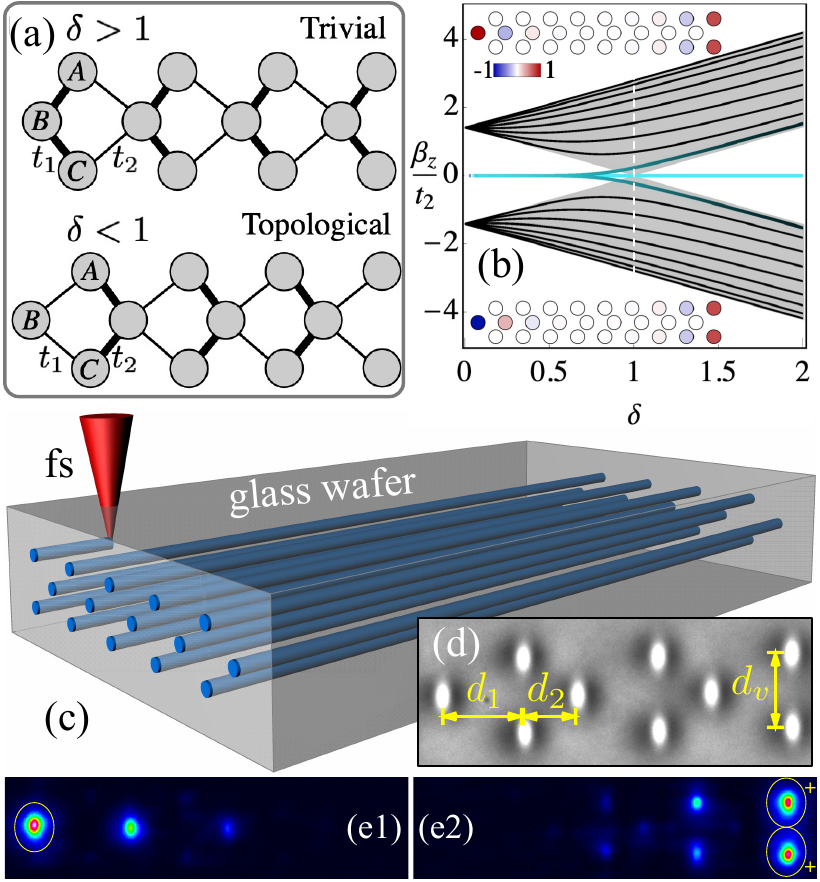}
\caption{(a) Sketch of a dimerized diamond lattice, with \textit{A}, \textit{B} and \textit{C} the sites of the unit cell. Thick (thin) line denotes a strong (weak) hopping, and $t_1$ ($t_2$) indicates the intra(inter)-cell coupling constant. Top (bottom) panel schematizes the trivial (topological) case $t_1>t_2$ ($t_1<t_2$). (b)  Spectrum as a function of $\delta$ for a finite (lines) and an infinite (shaded area) lattice. The vertical line denotes $\delta=1$. The color indicates the IPR for all the states. Inset: amplitude profiles of edge states at $\delta=0.4$. (c) Sketch of the fs laser writing technique. (d) Microscope image of a diamond lattice with $\{d_1,d_2\}=\{35,25\}\ \mu$m ($\delta=0.37$) and $d_v=32\,\mu$m. Output images for the lattice in (d) and for the excitation at (e1) a $B$ left edge site and (e2) an in-phase $A$-$C$ right edge sites. Yellow ellipses indicate the excited sites.}
\label{fig1}
\end{figure}

To demonstrate the topological edge transfer we use a diamond lattice of coupled waveguides with different intracell ($t_1$) and intercell ($t_2$) hoppings, as sketched in Fig.~\ref{fig1}(a). The lattice has three sites per unit cell, denoted as \textit{A}, \textit{B} and \textit{C} sites. Considering a tight-binding coupled-mode approach, the evolution of the optical field at every site of the n-th unit cell is written as:
%
\begin{align} 
\label{eq1}
-i\partial_zA_n &= t_1B_n+ t_2B_{n+1}\,,\nonumber\\ 
-i\partial_zB_n &= t_1(A_n+C_n)+ t_2(A_{n-1}+C_{n-1})\,,\\
-i\partial_zC_n &= t_1B_n+ t_2B_{n+1}\,\nonumber.
\end{align}
Here $A_n$, $B_n$ and $C_n$ are the amplitudes of the optical field at the n-th unit cell.
$z$ describes the coordinate along the waveguides and the dynamical variable. Moreover, the hopping strengths among nearest-neighbor (NN) sites can be varied experimentally by adjusting the lattice distances~\cite{Szameit2005}. We then define the control parameter $\delta\equiv t_1/t_2$ to characterize the different regimes. We assume an infinite system and impose a Bloch-like \textit{ansatz} in~\eq{eq1}, obtaining the following bands:
\begin{equation}
\label{bands}
\beta_z(k_x)=0,\pm t_2\sqrt{2\left[\delta^2+2\delta\cos(k_xa)+1\right]}\ ,
\end{equation}
where $\beta_z$ is the propagation constant (energy), $a$ is the lattice constant and $k_x$ the quasimomentum. The spectrum is composed of two dispersive and one flat band (FB) [see shaded areas and horizontal light-blue line at $\beta_z=0$ in Fig.~\ref{fig1}(b), respectively]. The gap in between both dispersive bands has a size equal to $2\sqrt{2}t_2|\delta-1|$. For $\delta=1$, this gap closes and the three bands touch each other at the edges of the Brillouin zone~\cite{Mukherjee2017}. The diamond lattice possesses the smallest experimentally reported FB states~\cite{Vicencio2021,Mukherjee2017}, with an inverse participation ratio (IPR)~\cite{StubPRR} of $1/2$ [represented as light-blue color in Fig.~\ref{fig1}(b)]. Specifically, in the bases of Wannier functions in the \textit{A}, \textit{B} and \textit{C} sites, the FB eigenvector is given by: $\ket{v^{FB}}=\{1,0,-1\}/\sqrt{2}$, and the ones corresponding to the dispersive bands are $\ket{v^\pm}=\{e^{i\phi(k_x)},\pm\sqrt{2},e^{i\phi(k_x)}\}/\sqrt{2}$, where $\phi(k_x)=\arctan(-\sin(k_xa)/[\delta +\cos(k_xa)])$. 

Even though this lattice has three sites per unit cell, it exhibits similar topological features to the SSH model~\cite{SSH1979}, when varying the parameter $\delta$~\cite{Bercioux2017}. Indeed, a quantized Zak phase of a value $0$ or $\pi$ can be found when $\delta>1$ ($t_1>t_2$) or $\delta<1$ ($t_1<t_2$), respectively. In this case, the nontrivial phase is protected by inversion symmetry between $A_n$ and $C_n$ and by chiral symmetry~\cite{Ramachandran2017,Bercioux2017}. Thus, we expect the appearance of two edge states at zero propagation constant on a lattice with open boundaries when $\delta<1$. To corroborate this, we compute the spectrum as a function of $\delta$ for dimerized diamond lattices of $9$ unit cells [see full lines in Fig.~\ref{fig1}(b)]. It can be clearly seen that two states at zero frequency (lighter blue) transform into two dispersive states (darker blue) around $\delta=1$. When increasing $\delta$, the degeneracy between them is removed at around $\delta=0.7$ (splitting $\Delta \beta_z^e\sim0.06$)~\cite{SM} due to the finite size of the lattice. The flat band remains unchanged at $\beta_z=0$, for any value of $\delta$. The IPR (denoted by color) shows very clearly the transition from localized edge states (IPR $=1$ or $1/2$, light blue) into extended propagating modes (IPR $\sim 1/N$, black).

Figs.~\ref{fig1}(b)-insets show the two edge states for $\delta=0.4$. They exhibit exponentially localized amplitudes at both edges. On the left edge, these states present a null amplitude at \textit{A} and \textit{C} sites, whereas the states have a null amplitude at \textit{B} sites at the right edge. Moreover, one edge state is antisymmetric (bottom inset) and the other one is symmetric (top inset), with respect to the opposite edge. They decay exponentially into the bulk as $(-\delta)^{|n-n_{e}|}$ for a semi-infinite system, exhibiting a phase shift of $\pi$ at consecutive \textit{B} or \textit{A,C} sites, depending on the specific edge ($n_e$). Notice in Fig.~\ref{fig1}(b) that the edge states are degenerate for $\delta\lesssim0.7$; consequently, the sum of these states gives a state fully localized at the left edge with amplitude on \textit{B} sites only and, conversely, the subtraction of them gives a state fully localized at the right edge with amplitude on \textit{A} and \textit{C} sublattices. For $\delta\gtrsim0.7$, the degeneracy of the edge states is lifted, and their frequency deviates from $0$ to $\pm \beta_z^e$. Therefore, the excitation of sites at the edges is expected to induce an oscillatory pattern in between both surfaces with a frequency $\beta_z^e$~\cite{Efre21}, with a long-distance state transfer occurring on a dynamical scale $z_{transfer}=\pi/\beta_z^e$~\cite{SM}. 

We fabricate several dimerized diamond photonic lattices, of $9$ unit cells each, by using a femtosecond (fs) laser writing technique~\cite{Davis96,Szameit2005,SM}, as it is sketched in Fig.~\ref{fig1}(c). 
For a first set of experiments, the diamond geometry is defined by distances $d_1$, $d_2$ and $d_v=32\ \mu$m, as described in Fig.~\ref{fig1}(d). For these values, the diagonal (NN) distance was swept in the interval $\{25.6,43.1\}\ \mu$m, as $d_1$ and $d_2$ were varied in the interval $\{20,40\}\ \mu$m in steps of $1\,\mu$m. The hopping coefficients (which decay exponentially on waveguide separation~\cite{Szameit2005}) range in the interval $\sim\{0.03,0.21\}$ cm$^{-1}$ at a wavelength of $640$ nm~\cite{SM}. Fig.~\ref{fig1}(d) shows an output facet of a lattice with $d_1=35$ and $d_2=25\ \mu$m, with $t_1=0.05$ and $t_2=0.14$ cm$^{-1}$ ($\delta=0.37$). We first test the quality of the lattices by exciting them at different input positions using a $640$ nm laser beam (see Ref.~\cite{SM} for a complete characterization). For example, topological edge states can be efficiently excited by injecting light directly at the lattice boundaries~\cite{Malkova2009,StubPRR,Wang2022,SM}. Fig.~\ref{fig1}(e1) shows the output profile after a \textit{B}-edge site excitation, with a clear exponential decaying profile from the edge into the bulk. The excitation at the right boundary requires a more complicated input condition with two in-phase beams. The result of this is shown in Fig.~\ref{fig1}(e2), with an output profile formed by \textit{A} and \textit{C} sites mostly.



\begin{figure}[t!]
\centering
\includegraphics[width=\columnwidth]{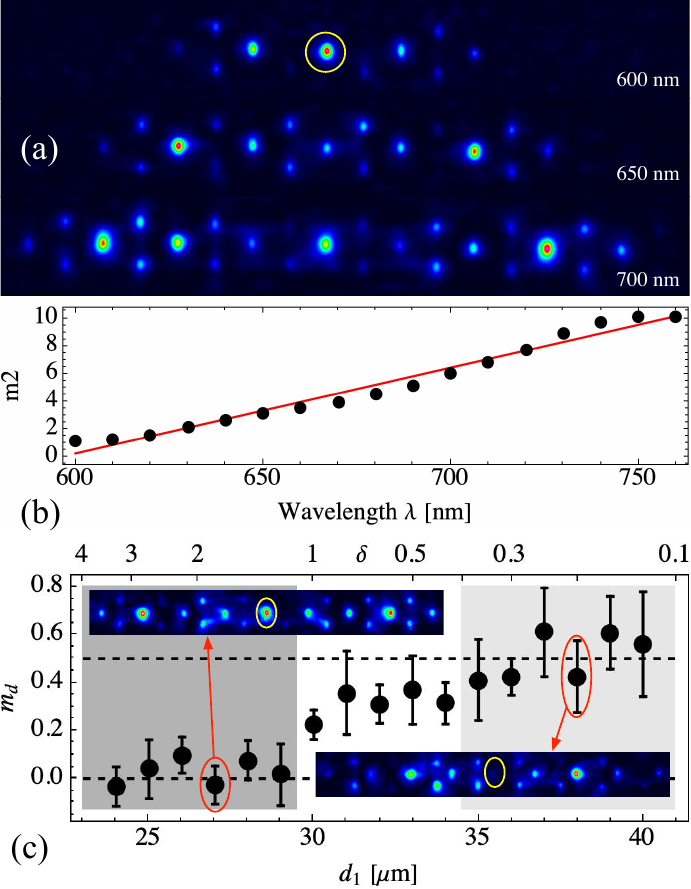}
\caption{(a) Output intensity profiles for a \textit{B}-site central excitation, for a lattice with $d_1=d_2=30\ \mu$m, at the indicated wavelength. (b) $m2$ versus $\lambda$. (c) $m_d$ versus $d_1$ (bottom) and $\delta$ (top). Insets in (c) show the profile at $700$ nm for the indicated case. The bar shows the standard deviation. Yellow ellipses indicate the excited sites.}
\label{fig2}
\end{figure}

We propose a novel technique to characterize the lattice dynamics. Instead of measuring the output profiles at different $z$ values (which also implies the fabrication of a larger number of lattices), we implement a \textit{wavelength-scan method}: The dynamics of a wavepacket injected in the input facet of a lattice is revealed when varying the input wavelength coming from a Supercontinuum (SC) laser source. In general, the lattice dynamics depends on the excitation wavelength $\lambda$: the longer the wavelength the wider the mode profile and the larger the coupling constants~\cite{Szameit2005,SM}. In this way, by tuning the input wavelength, we can study the same lattice at different effective lengths. 

We first consider a diamond lattice with $d_1=d_2=30\ \mu$m. We excite a \textit{B} site at the central $5$-th unit cell and scan the input wavelength in the interval $600-760$~nm, with a step of $10$ nm. Fig.~\ref{fig2}(a) shows the output intensity for three selected $\lambda$'s~\cite{SM}. Fig.~\ref{fig2}(b) shows the \textit{second moment (width)}, defined as $m2\equiv\sum_n (n-n_c)^2P_n$, versus wavelength, with $P_n\equiv|A_n|^2+|B_n|^2+|C_n|^2$ the unit cell power and $n_c\equiv\sum_n n P_n$ the \textit{center of mass}. We observe a growing diffraction pattern~\cite{Vicencio2021}, with a width that increases almost linearly with the input wavelength [a linear fit is included in Fig.~\ref{fig2}(b)]. $m2\sim z$ corresponds to a diffusive regime~\cite{Naether_2013}, as expected for discrete diffraction; therefore, a $\lambda$ increment produces an effectively larger propagation distance $z$ or a larger coupling constant $t_{1,2}$.

A dimerized diamond lattice has two hoppings which simultaneously change while $\lambda$ is modified. Since we observe a linear dependence of coupling constants on wavelength, we can assume $\delta$ as a constant, as a first approximation. We use the wavelength-scan method to experimentally determine $n_c$ for all the output profiles, after exciting a \textit{B} site at the central ($5$-th) unit cell of $17$ dimerized lattices, having different values of $\delta$. For each lattice, we average $n_c$ over $\lambda$ and obtain the \textit{averaged beam displacement} $m_d$, from which the topological invariant can be inferred~\cite{Longhi2018,LonghiMean21}. A topologically trivial lattice has a $m_d= 0$, as an indication of a zero Zak phase. A topologically non-trivial system will shift this value to $m_d\sim0.5$, corresponding to a $\pi$ Zak phase~\cite{Longhi2018}. Our collected results are shown in Fig.~\ref{fig2}(c). We observe that for $d_1<30\ \mu$m ($\delta>1$) the lattice is topologically trivial and the propagation shows a $m_d$ around zero. For $30\leqslant d_1\leqslant34\ \mu$m ($0.4<\delta\leqslant1$), a transition region without a well defined topological phase is observed. For $d_1\geqslant35\ \mu$m ($\delta\leqslant0.4$), the lattices express a clear averaged beam displacement around $0.5$, implying a nontrivial Zak phase. Therefore, the wavelength-scan method gives us valuable information about the dynamics on a specific lattice, and it becomes a key method to determine its topological phase on a finite size configuration.

\begin{figure}[tbp!]
\centering
\includegraphics[width=0.99\columnwidth]{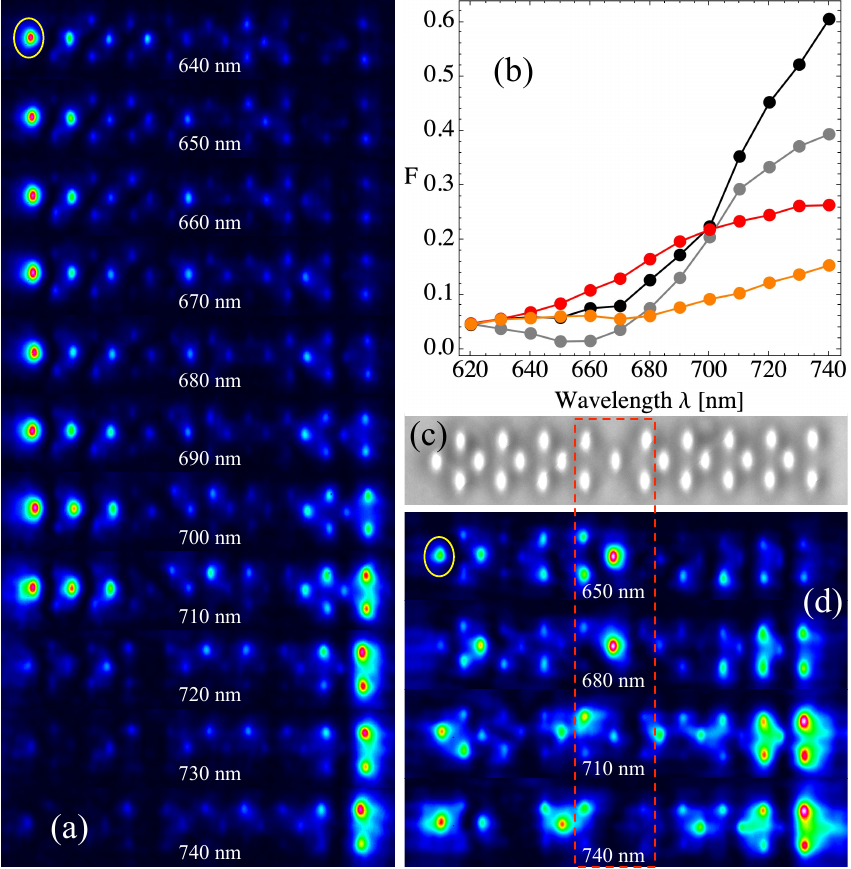}
\caption{(a) Output profiles of a non-trivial diamond photonic lattice at different $\lambda$'s, after a \textit{B}-edge excitation (see yellow ellipse). (b) Fidelity versus wavelength for topological (black), trivial (gray), topological + defect (red), and trivial + defect (orange) lattices. (c) Microscope image of a topological lattice plus two coupling defects (see dashed rectangle). (d) Same than (a) for a topological lattice with a coupling defect.}
\label{fig3}
\end{figure}
%
The number of unit cells in the lattice affects the edge state properties: the fewer unit cells, the shorter the range of $\delta$ in which the edge states keep degenerate in $\beta_z$~\cite{SM}. When the degeneracy is lifted, the two edge modes hybridize. 
Therefore, an input on one edge will excite coherently both modes and result in periodic oscillations of the amplitude at the two edges. Then, a transfer of light from one edge to the other becomes possible~\cite{Efre21,DeLeseleuc2019}. To experimentally demonstrate this we fabricate a topological lattice with $9$ unit cells and distances $d_1=18$, $d_2=14$, and $d_v=14\ \mu$m ($t_1=0.30$ and $t_2=0.42$~cm$^{-1}$ at $640$~nm, and $\delta=0.71$). The trivial lattice ($\delta=1.40$) is obtained by inverting these distances to $d_1=14$ and $d_2=18\ \mu$m. We decreased the distances to increase the coupling coefficients and favor a faster transport in between the edges, while staying at the non-degenerate situation. Again, we use a SC laser source in the range $610-740$ nm and sweep the input wavelength in steps of $10$ nm. We excite the system by injecting light at the \textit{B} left edge waveguide, as shown in Fig.~\ref{fig3}(a). For $\lambda\lesssim670$~nm, the intensity profiles are well localized at the left edge, with most of the light intensity at the \textit{B} sublattices, with a profile resembling the edge state [Fig.~\ref{fig1}(e1)]. The edge states splitting manifests for $\lambda\approx680$~nm, where we start observing a smooth population of the opposite edge, with a weak excitation of the lattice center (a weak background radiation is always observed because of the excitation of dispersive modes~\cite{SM}). The connection in between both edge patterns [see Figs.~\ref{fig1}(b)-insets and (e)], with a \textit{B}-site exponential decaying profile at the left surface and an \textit{A,C} exponential profile at the right edge, becomes evident for $\sim710$~nm. The spectral state transfer phenomenon starts occurring at $\lambda\gtrsim720$~nm: the light injected at one edge is mostly transferred into the opposite edge~\cite{SM}. This shows a very interesting transport mechanism which does not require that the light explores the whole lattice; in this case, the light is suddenly transferred from one edge into the other without interacting with the lattice bulk. 

We define the fidelity $F$ for an edge-to-edge light transfer by measuring the normalized transferred intensity at the opposite lattice edge: $F\equiv(|A_{edge}|^2+|C_{edge}|^2)/\sum_nP_n$. If all the light reaches the two rightmost sites $F=1$, and $F=0$ in the fully opposite case. We show our results in Fig.~\ref{fig3}(b), where we plot the fidelity $F$ versus $\lambda$, for topological and trivial lattices. We observe how the topological (black) and the trivial (gray) cases have a similar dynamical scale; i.e., both processes occur approximately at the same speed. However, the fidelity at the $A,C$ surface is larger for the topological lattice ($\sim61\%$). The trivial lattice presents a standard discrete diffraction pattern~\cite{Vicencio2021}, with the energy exploring the whole lattice while it moves from one edge into the other~\cite{SM}, as the wavelength increases [similar to Fig.~\ref{fig2}(a)]. Therefore, once the light arrives at the $A,C$ right edge it is reflected back due to the absence of the edge states. The fidelity in this case decreases to a $\sim40\%$. 

A remarkable feature of the state transfer between topological edge states is the resilience to certain types of disorder. Although the fabrication process can produce random on-site or inter-site defects, we fabricate a couple of lattices with a symmetric coupling defect, as the one shown in Fig.~\ref{fig3}(c). We design a different distance in between the fourth and the fifth cells and inside the fifth cell [see dashed rectangle in Fig.~\ref{fig3}(c)]. Specifically, we set this distance to $23\ \mu$m, implying a coupling defect of $0.18$~cm$^{-1}$. Fig.~\ref{fig3}(d) shows a set of output images at the indicated values of $\lambda$, for the topological lattice with a defect. We notice that this defect produces some reflection and trapping of energy at short wavelengths, consequently, not all the energy is edge-to-edge transferred. Despite this, a significant amount of the light excites the topological right-edge state composed of $A$ and $C$ sites. The fidelity is $\sim26\%$ for the topological case, whereas it drops to $\sim15\%$ for the trivial lattice. 

These numbers show that a trivial lattice undergoes a stronger back reflection caused by the defect, because the light explores the whole lattice and interacts strongly with it. On the other hand, in the topological case, the light does not travel across the lattice and excite efficiently the edge state without the need of arriving at the boundary by standard transportation. The fidelity is not perfect in none of the topological cases because a single $B$-site input always excites part of the dispersive spectrum, in which the modes extend over the entire lattice. Nonetheless, the strong difference between the topological and trivial cases is the key of success for a topological state transfer process, which occurs due to the excitation of exponentially localized edge states which live at both edges simultaneously and deep in the gap of the lattice spectrum. This could be a proof of concept for a long distance sensor, which detects away from the interaction region. 


The wavelength-scan method proposed in this work offers a tool for investigating the dynamics in lattices of coupled waveguides. Using this method, we evidenced the nontrivial topology of dimerized diamond lattices by measuring the averaged beam displacement and, additionally, we demonstrated an edge-to-edge transfer of light via the excitation of the topological edge states. This transfer is partially robust to defects across the lattice bulk and, hence, it has the potential for a precise wavelength filter, as well as an efficient information transport mechanism between two distant ports or by concatenating several topological lattices in the quantum domain~\cite{Lang2017,Mei2018}.


\textit{Acknowledgments}. This work was supported in part by Millennium Science Initiative Program ICN17$_-$012, and FONDECYT Grant 1191205. A.A. acknowledges the support of European Research Council grant EmergenTopo (865151), the French government through the Programme Investissement d’Avenir (I-SITE ULNE / ANR-16-IDEX-0004 ULNE) managed by the Agence Nationale de la Recherche, the Labex CEMPI (ANR-11-LABX-0007) and the CPER Wavetech. T.O. acknowledges the support from JSPS KAKENHI Grant No. JP20H01845, JST PRESTO Grant No. JPMJPR19L2, and JST CREST Grant No.JPMJCR19T1.

$^{\dagger}$Both authors contributed equally.

\bibliography{biblio}{}

\end{document}